\documentclass[reprint,aps,prb,twocolumn,superscriptaddress,showpacs]{revtex4-1}
\usepackage{graphicx}
\usepackage{mathrsfs}
\usepackage{bm}
\usepackage{hyperref}
\usepackage{dcolumn}
\usepackage{amsmath}
\usepackage{mathtools}
\usepackage{amssymb}
\usepackage{color}
\usepackage[normalem]{ulem}
\usepackage{bbm}
\usepackage{bbold}

\hypersetup{  colorlinks=true, linkcolor=blue, citecolor=red, urlcolor=blue  }

\newcommand{\bs}{\bm{\mathrm{\sigma}}}
\newcommand{\bB}{{\mathbf B}}
\newcommand{\bk}{{\mathbf k}}
\newcommand{\bp}{{\mathbf p}}

\begin{document}
\title{Majorana Zero Modes in Cylindrical Semiconductor Quantum Wire}

\author{Chao Lei}
\affiliation{Department of Modern Physics, University of Science and Technology of China, Hefei, Anhui 230026, China}
\affiliation{Department of Physics, The University of Texas at Austin, Austin, Texas 78712,USA}
\author{Guru Khalsa}
\affiliation{Department of Physics, The University of Texas at Austin, Austin, Texas 78712,USA}
\affiliation{Department of Materials Science and Engineering, Cornell University, Ithaca, New York 14853, USA}
\author{Jiangfeng Du}
\affiliation{Department of Modern Physics, University of Science and Technology of China, Hefei, Anhui 230026, China}
\author{Allan H. MacDonald}
\affiliation{Department of Physics, The University of Texas at Austin, Austin, Texas 78712,USA}

\begin{abstract}
We study Majorana zero modes properties in cylindrical cross-section semiconductor quantum wires based on 
the $\bk \cdot \bp$ theory and a discretized lattice model.  Within this model the influence of disorder potentials in the wire 
and amplitude and phase fluctuations of the superconducting order-parameter are discussed. 
We find that for typical wire geometries, pairing potentials, and spin-orbit coupling strengths,
coupling between quasi-one-dimensional sub-bands is weak, low-energy quasiparticles near the Fermi energy are nearly completely spin-polarized, 
and the number of electrons in the active sub-bands of topological states is small. 
\end{abstract}

\maketitle

\section{Introduction}

One-dimensional (1D) p-wave superconductors are topologically nontrivial\cite{Kitaev} and, in finite systems, support 
end-localized Majorana zero modes.\cite{Majorana}
These states have attracted considerable interest lately\cite{ReviewDasSarma,ReviewAlicea,ReviewBeenakker,Elliott2015,ReviewLutchyn,
ReviewLeijnse,ReviewStanescu}
because of their non-Abelian exchange properties,\cite{Read,Ivanov} and related potential utility in quantum information processing systems\cite{ReviewDasSarma}.
Theory has suggested\cite{PriorTheory_1,PriorTheory_2} that it should be
possible to engineer effective one-dimensional p-wave superconductors in proximity coupled 
semiconductor quantum wires by combining broken inversion 
symmetry, and the consequent Rashba spin-orbit interactions, with external magnetic fields.  
Considerable progress has been made in exploring this idea experimentally.\cite{Expt_Delft,Expt_Xu,Expt_Weizmann,Expt_Marcus,Expt_Finck,Expt_Rokhinson,
albrecht2016,Deng2016,Exp_Kouwenhoven_2018,Chen2017,Suominen2017,Nichele2017,
Zhang2018,Sestoft2018,Deng2018,Vaitiek_nas_2020,Zhang2020,Zhang2020b,Li2020,Ekstrom2020,Lee_2019,Ricco2019,Winkler2019,Antipov2018,
2D_Chan2017,2D_kjaergaard2016,2D_Nichele2017,2D_bottcher2018,2D_Pientka2017,2D_Sjabamo2016,2D_Hell2017,2D_Suominen2017}
There has also been progress toward Majorana-based quantum state manipulation 
in other systems, including magnetic atom chains,\cite{Expt_Nadj,Pawlak2016,Ruby2015}
interfaces between conventional superconductors and topological insulators,\cite{Wang2012,He2017}
iron-based superconductors,\cite{Wang2018} and phase-controlled Josephson junctions.\cite{Yacoby1,Yacoby2} 

The Majorana zero modes in semiconductor quantum
wires\cite{Majorana_Early_1,Majorana_Early_2,Majorana_Early_3,Majorana_Early_4}
are expected to appear only when external magnetic field strengths exceed
a critical value, beyond which the proximity-induced superconductor gap vanishes.
Early experiments in cylindrical cross section quantum wires exhibit many
trends consistent with expectations\cite{Expt_Delft,Expt_Xu,Expt_Weizmann,Expt_Marcus,Expt_Finck,Expt_Rokhinson} 
based on Majorana zero mode properties,
although they also consistently exhibit evidence of a soft gap, {\it i.e.} of quasiparticle states within the 
gap, at all magnetic field strengths.
The in-gap states can be associated with spatially extended Andreev states,\cite{Andreev_Liu} disorder\cite{Disorder_Dmitry,Disorder_Liu,Disorder_Pikulin} or Kondo effects,\cite{Kondo_Lee} and may influence electron 
transport experiments, and would poison any attempt to achieve topologically protected state manipulation.

%Researchers have attempted\cite{Hard_gap_2015,Hard_gap_2017,krogstrup2015}  to make devices with hard superconducting gaps by growing heterojunctions with cleaner interfaces between semiconductor quantum wires and superconducting metals,
%for example by growing hexagonal (Hex in Figure~\ref{cross_section}) cross section semiconductor  quantum wires\cite{albrecht2016,Albrecht2017,Deng2016,Exp_Kouwenhoven_2018,Chen2017,Suominen2017,Nichele2017,Zhang2018,Sestoft2018,Deng2018} 
%surrounded by deposited superconducting metal.
%In addition to understanding fundamental properties of engineered Majorana zero modes, the goal of developing a scalable architecture for Majorana-based quantum computing, by fabricating quantum wire T-junction\cite{Alicea2011} networks in two-dimensional planar structures \cite{2D_Chan2017,2D_kjaergaard2016,2D_Nichele2017,2D_bottcher2018,2D_Pientka2017,2D_Sjabamo2016,2D_Hell2017,2D_Suominen2017}
%defined by top-down (lithography-based) technologies remains of high interest but comes with the hurdle of so-called bend-potential effects (add references) -- bound states at the junction between two wires -- which may stifle topologically protected manipulation of Majorana zero modes.

In this paper, we study quasi-one-dimensional cylindrical quantum wires numerically, using experimentally realistic geometries 
diameters $\sim 100$ nm, as shown in Fig. \ref{cross_section}), experimentally estimated pairing potential and spin-orbit coupling strengths, 
and a variety of types of experimentally realistic disorder.  In experiment, the longest quantum wires have approximate cylindrical cross-sections.
Longer wires have weaker hybridization between Majorana zero modes at the ends of quantum wire, 
and more electrons in active subbands.

\ifpdf
\begin{figure}[htbp]
\centering
\includegraphics[width= 1\columnwidth]{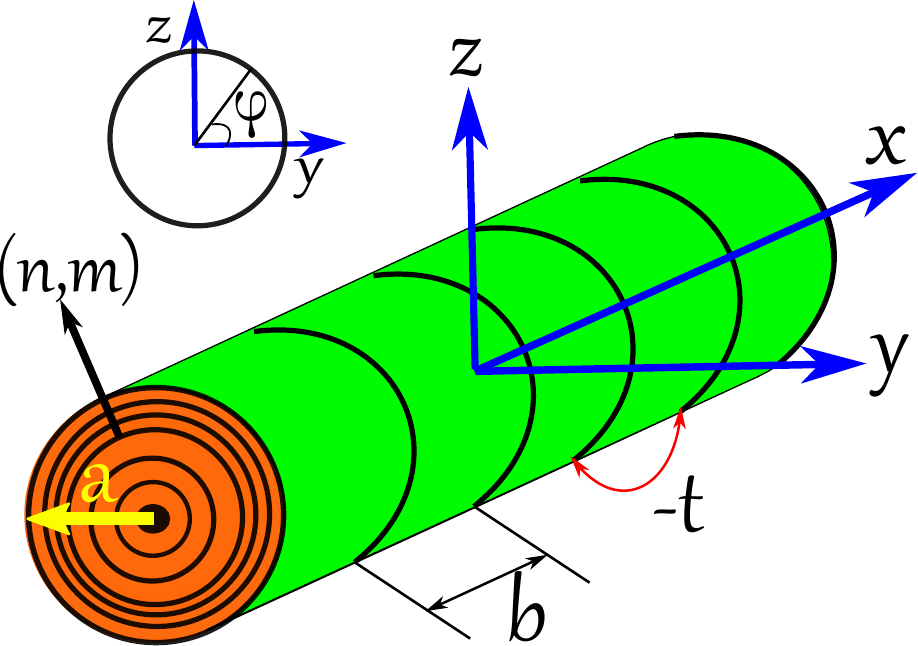}
\caption{(Color online).
Cylindrical semiconductor quantum wire geometry. Here $a$ labels the radius of the cylindrical quantum wire, $b$ is the lattice constant
used to discretize position along the wire in our numerical studies, and $t$ the corresponding hopping strength. 
The circles in the cross-section schematically represent radial wavefunctions labeled by principal axial quantum number $n$ and angular momentum $m$.
}
\label{cross_section}
\end{figure}
\fi

Our paper is organized as follows.
In Section \ref{cylindrical_cross_section} we introduce a theoretical model for cylindrical quantum wires 
and discuss its topological-state phase diagram as a function of Fermi energy and magnetic field.
In Section \ref{andreev_states} we analyze the Andreev states, and the tunneling density of states as a function of 
magnetic field, disorder potential, and pairing-potential disorder in infinite quantum wires.
In Section \ref{majorana_states_properties} we address the case of finite length, using wire lengths on the scale of 
experimental samples and discuss finite-length Majorana energy splitting effects.
In Section \ref{discussions} we discuss the use of models in which only degrees of 
freedom in the semiconductor quantum wire are included explicitly, {\it vs.} models that account explicitly
for the superconducting metal.

\section{$k \cdot p$ theory}\label{cylindrical_cross_section}

When Rashba spin-orbit interactions are neglected, the cylindrical-coordinate $\bm{k} \cdot \bm{p}$ Hamiltonian for
an $n$-type semiconductor quantum wire oriented along the $\hat{x}$ direction(shown in Fig. \ref{cross_section})
separates into a free-particle contribution along the wire and a radial confinement contribution.\cite{Dresselhaus,SooLim_2013}  The Hamiltonian is
\begin{equation}\label{h0}
    {\mathcal{H}_0} = \frac{{{\hbar ^2}}}{{2{m^*}}}(k_x^2 - \frac{{{\partial ^2}}}{{\partial {r^2}}} - \frac{1}{r}\frac{\partial }{{\partial r}} - \frac{1}{{{r^2}}}\frac{{{\partial ^2}}}{{\partial {\varphi ^2}}}) + V(\textbf{r},x)
\end{equation}
where $ \hbar $ is Planck's constant, $ m^* $ is the conduction band effective mass, $ V(\textbf{r},x) $ is the confining
potential, $\textbf{r} = (y,z)=(r\cos(\varphi),r\sin(\varphi))$ is the position projected to the
wire cross-section, $ x $ is position along the wire, and $k_x$ is wave vector along the wire.
In the absence of disorder, we take $V(\textbf{r},x) $ to be $ 0 $ inside the wire ($ \left| \textbf{r} \right| < a $ where
$ a $ the radius of the wire) and $+ \infty$ outside the wire.

Cylindrical symmetry implies that eigenstates can be labeled by angular momentum $m$ along the wire axis.
The confined radial wave functions are then Bessel functions with zeros at the wire edge. The one-dimensional transverse wave-functions are
\begin{equation}\label{function}
   f_{n,m} (r,\varphi ) = \begin{array}{*{20}{c}}
  {A_{n,m}{J_{|m|}}({u_{n,m}}\frac{r}{a}) \, {e^{im\varphi }},} \; &{m = 0, \pm 1, \pm 2,...}
\end{array}
\end{equation}
where $ {J_{|m|}({u_{n,m}}\frac{r}{a})} $ is an $ m^{th}-$order Bessel function,
$ u_{n,m} $ is the $ n^{th}$ zero of the $ m^{th}$-order Bessel function, and
$ {A_{n,m} = 1/[a\sqrt \pi  {J_{|m|+1}}({u_{n,m}})]} $ is a normalization constant.
%{\bf Chao: Is the preceding equation really correct?  Can you remind me how to derive it?  Perhaps this is the
%expression for $A^2$.  Or is the Bessel function supposed to be under the sqrt.  Please correct.}
The one-dimensional sub-bands are rigidly offset by an energy which is determined by the principal axial quantum number $n$ and the azimuthal quantum number $m$ that quantifies the angular momentum, the dispersion is
\begin{equation}
E_{n,m}(k_x)  = \frac{{{\hbar ^2}}}{{2{m^*}}}k_x^2 + \frac{{{\hbar ^2}}}{{2{m^*}}}\frac{{u_{n,m}^2}}{{{a^2}}}.
\end{equation}
\noindent
Note that since $u_{n,m}=u_{n,-m}$ so $|m| \ne 0$ sub-bands are always doubly degenerate.

The one-dimensional band structure with quantum numbers labeled by (n,m) is illustrated in Fig.~\ref{Bandstructure_cyl}. These results were obtained by using parameters that are appropriate for the $ a = 50 $ nm InSb quantum wire ($m^* = 0.015 m_e $) studied in the first Majorana experiment \cite{Expt_Delft} with Rashba coupling parameter $ \alpha$ = 0.02 eV $\cdot$ nm.
We note that subsequent experiments studied quantum wires with similar properties.
%Note that the lowest energy sub-band has $m=0$, that the 2$^{nd}$ and 3$^{rd}$ sub-bands and 5$^{th}$ and 6$^{th}$ sub-bands have $m = \pm 1$, and that the fourth sub-band again has $m=0$.
Rashba spin-orbit interactions lift the $m = \pm 1$ degeneracy and for finite $k_x$
and lift the spin-degeneracy within each sub-band.

\ifpdf
\begin{figure}[htbp]
\centering
\includegraphics[width= 1\columnwidth]{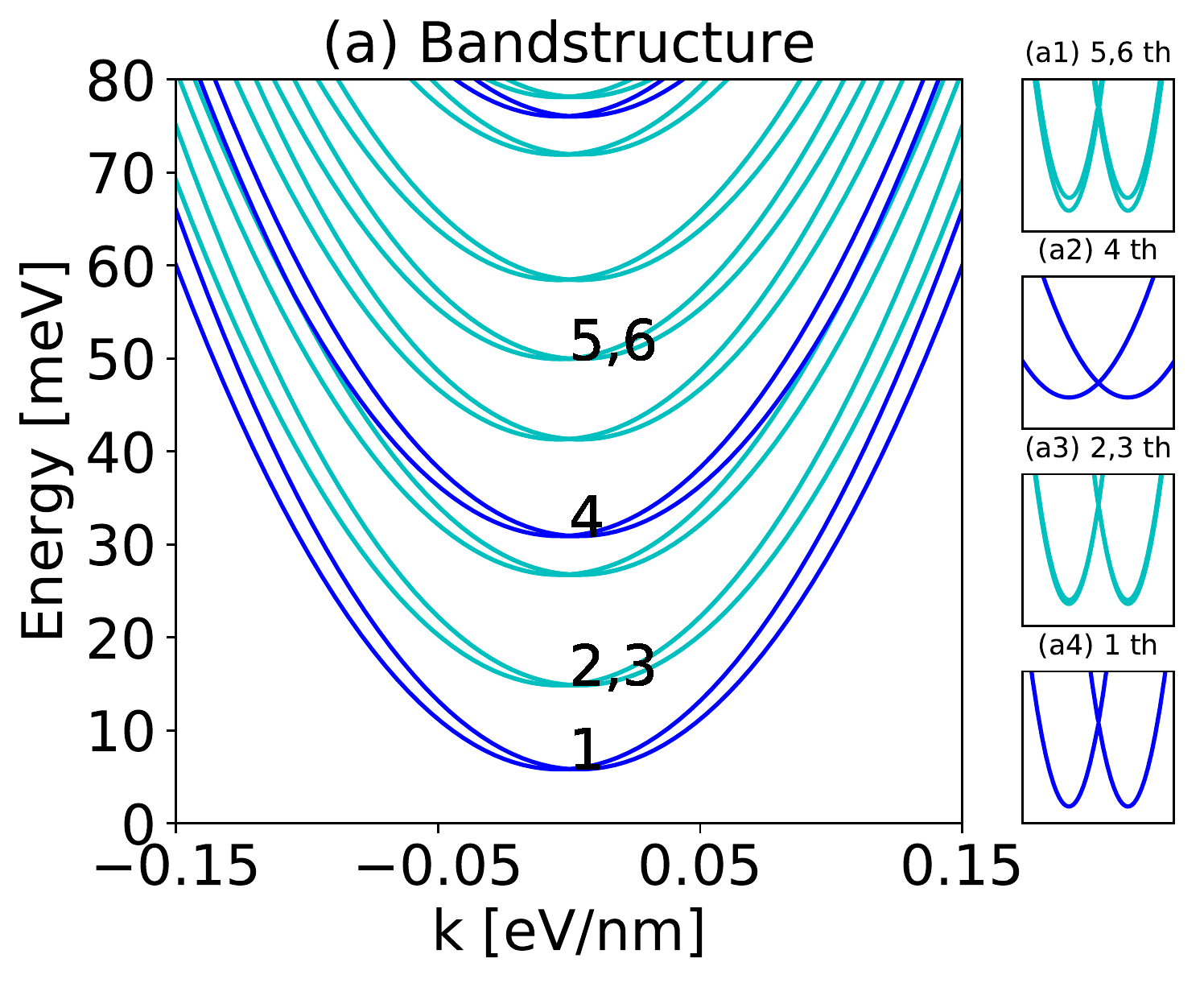}
\caption{(Color online).
One-dimensional band-structure of a cylindrical semiconductor quantum wire with radius $ a = 50 $nm, InSb conduction band mass $m^* = 0.015 m_e $, and Rashba coupling parameter $ \alpha$ = 0.02 eV $\cdot$ nm. The panels on the right highlight the behaviors near band minima, which are important for topological superconductivity. Angular momentum $m$ is not a good quantum number for finite Rashba coupling strength. Because the Rashba interaction couples only states that differ by $\pm 1$ in angular momentum. The mixing between $m=1$ and $m=-1$ sub-bands is second-order in the ratio of the Rashba coupling strength ($\sim \alpha/a$) to the sub-band separations, which is small.
}
\label{Bandstructure_cyl}
\end{figure}
\fi

The mean-field Hamiltonian of a spin-orbit coupled quantum wire with proximity-induced $s$-wave superconductivity
and an external magnetic field includes one-dimensional sub-band, Rashba, Zeeman, and pairing contributions:
\begin{equation}\label{ham}
    \mathcal{H} = \mathcal{H}_0 + \mathcal{H}_R +  \mathcal{H}_Z + \mathcal{H}_{SC}
\end{equation}

\noindent
It is convenient to express this Hamiltonian in the representation of parabolic band quantum wire eigenstates. Assuming that the quantum wire is placed on a
substrate with a $\hat{z}$ direction surface normal, the quantum wire Rashba Hamiltonian  is
\begin{equation}
{\mathcal{H}_R} = \alpha \left[ { - i(\cos \varphi \frac{\partial }{{\partial r}} - \frac{{\sin \varphi }}{r}\frac{\partial }{{\partial \varphi }}){\sigma _x} - {k_x}{\sigma _y}} \right],
\end{equation}

\noindent
where $ \alpha $ is the Rashba coupling parameter and $\sigma_{\alpha}$ is a
Pauli matrix acting on spin. The matrix elements of the Rashba Hamiltonian in the representation of unperturbed band states are
\begin{equation}
\langle n,m| \mathcal{H}_R | n',m' \rangle =   - \alpha {k_x}{\sigma _y}\delta _{n,n'}{\delta _{m,m'}} -
i\alpha {R_{nm;n'm'}}\sigma _x,
\end{equation}
\noindent
where
\begin{equation}
R_{nmn'm'} = \left\langle {{f_{nm}}} \right|(\cos \theta \frac{\partial }{{\partial r}} - \frac{{\sin \theta }}{r}\frac{\partial }{{\partial \theta }})\left| {{f_{n'm'}}} \right\rangle
\end{equation}

\noindent 
is non-zero for $m= m' \pm 1$. 

The Zeeman Hamiltonian can be written as: $ \mathcal{H}_Z = \bB  \cdot \bs $
where $ \bB  $ is the magnetic field expressed in energy units.
In most experiments the magnetic field is along the
$ \hat{x} $ direction.  %{\bf Chao: The draft said x-direction.  Isn't it z-direction?}
The proximity-induced s-wave pairing contribution to the Hamiltonian is
\begin{equation}
\mathcal{H}_{SC} = \sum_{n,m} \big[ {\Delta_{SC}^*}c_{nm \downarrow k}^\dag c_{nm \uparrow  - k}^\dag  + \Delta_{SC} {c_{nm \uparrow  - k}}{c_{nm \downarrow k}}\big] ,
\end{equation}
\noindent
where $ \Delta_{SC} = |\Delta_{SC}|e^{i\phi} $ is the proximity induced gap.  The value of $\Delta_{SC}$ depends on a complex hybridization processes between orbitals in the quantum wire and orbitals in the surrounding superconductor but can be fit to experimental observations.  The relatively large values of $\Delta_{SC}$ ($0.25$ meV in Ref. \onlinecite{Expt_Delft} for example) suggest that
the interface between the quantum wire and the surrounding superconductor is quite transparent. We will return to his point in the discussion section.

Topologically distinct phases are separated in coupling-constant parameter space by gapless boundary states.  In the case of topological superconductivity in quantum wires, the coupling constants that are readily varied in experiments are the position of the Fermi level relative to the conduction band minimum, which can be altered by manipulating gate voltages, and the strength of the magnetic field responsible for Zeeman coupling to the spin degree of freedom. In the absence of an external magnetic field all states are topologically trivial. As the magnetic field strength is increased, the energy gap produced by the proximity effect pairing potential sometimes closes at discrete points.  The phase diagram in Fig.\ref{phase_diagram} was constructed by tracking these band closings and identifying each with a phase transition from a topologically trivial to a nontrivial state. While increasing the magnetic field with the Fermi level positioned near a sub-band, the superconductor gap closes when the system is driven from trivial superconductivity to non-trivial topological superconductivity. The phase diagram can be found by tuning the Fermi level and magnetic field (shown as Fig. \ref{phase_diagram}). When the Zeeman Energy exceed
the pairing potential while the Fermi level is tuned to lie at the bottom of a sub-bands a Majorana(-like) zero-mode phase appears. For sub-bands $n = 1,2$ and $m = 0,\pm1$ (see Fig.\ref{phase_diagram}), these six sub-bands have the Fermi energy of $ E_{F1} \approx 5.87 $ meV, $ E_{F2} \approx E_{F3} \approx 14.9 $ meV, $ E_{F4} \approx 30.96 $ meV and $ E_{F5} \approx E_{F6} \approx 50 $ meV. In the phase diagram we use Roman numerals to label the number of Majorana-like end-localized states. ($I$ -- one localized state, $II$ -- two localized states, etc.) The bottom panel of Fig. \ref{phase_diagram} shows the phase diagram for small Zeeman energy where we note that more detail can be seen. Due to the lack of degeneracy of the sub-bands in Fig. \ref{phase_diagram}(b,d) we expect robust Majorana modes, but in Fig. \ref{phase_diagram}(c,e) where sub=sbands are neirly degenerate, we expect Majorana-like modes that are weakly coupled.

\ifpdf
\begin{figure}[htbp]
\centering
\includegraphics[width= 1\columnwidth]{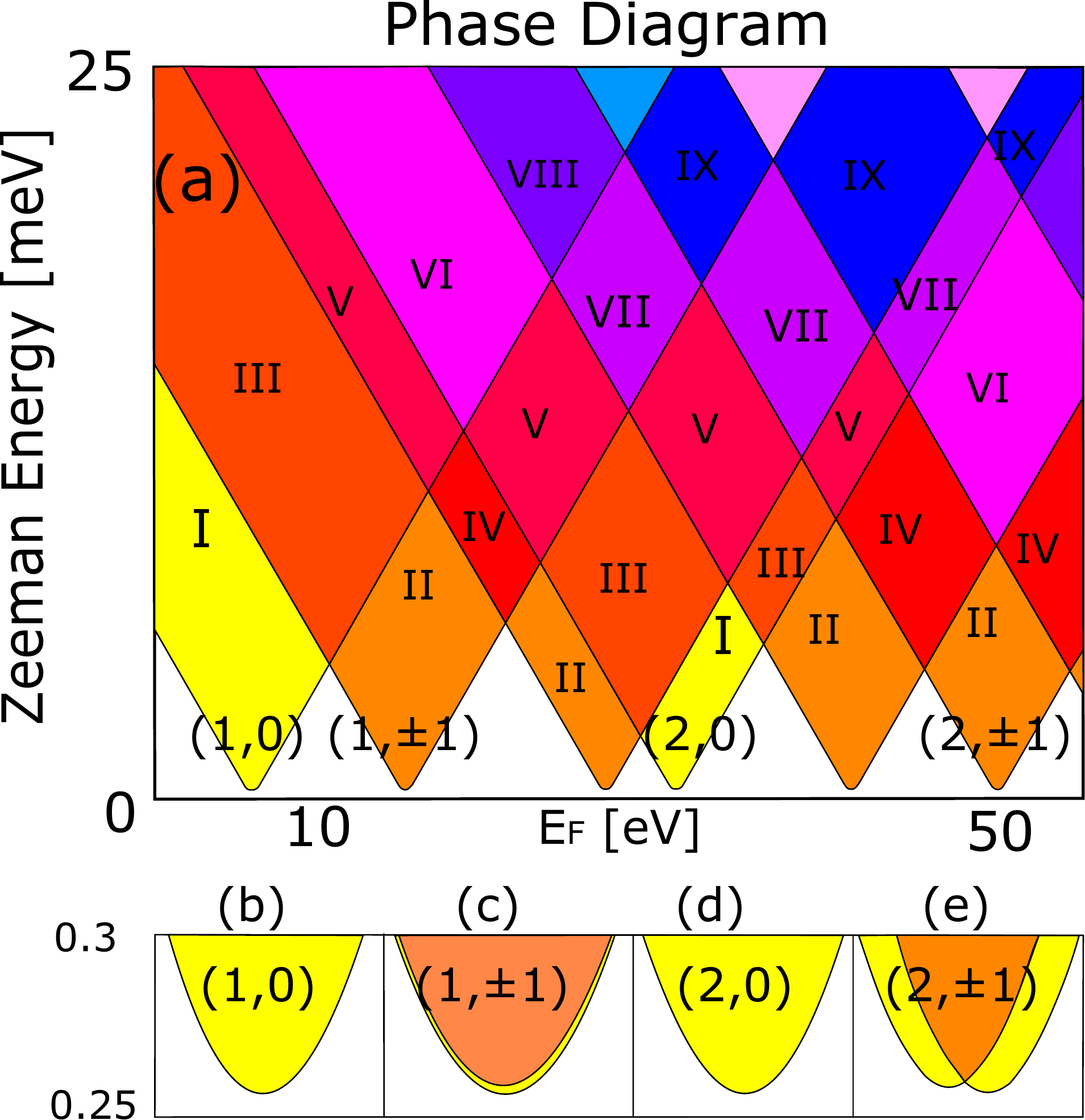}
\caption{(Color online).
   Phase diagram of the cylindrical semiconductor quantum wire as a function of Fermi level and magnetic field. (a) At each sub-band there appears Majorana(-like) zero-modes when the magnetic field exceeds the pairing potential and the Fermi level is tuned to the bottom of the sub-bands, of which n = 1,2 and m = $0,\pm1$ sub-bands have the Fermi energy at $ E_{F1} \approx 5.87 $ meV, $ E_{F2} \approx E_{F3} \approx 14.9 $ meV, $ E_{F4} \approx 30.96 $ meV and $ E_{F5} \approx E_{F6} \approx 50 $ meV. In the figure we use Roman numerals and color coordination to label the number of Majorana-like end-localized states. 
   (b-e) Phase diagram focused near the band minima for different (n,m).
}
\label{phase_diagram}
\end{figure}
\fi

\section{Andreev states}\label{andreev_states}

In contrast to the Majorana modes, zero-bias conductance peaks (ZBCP) in transport experiments may also come from Andreev states which was recently strudied experimentally.\cite{Deng2016} Here we distinguish the evolution of Andreev states from Majorana zero modes by varying the magnetic field. From the discussion of Section \ref{cylindrical_cross_section}, we see that there are degenerate sub-bands which are weakly coupled by Rashba interactions for non-zero angular momentum, while the zero-angular momentum sub-bands are not degenerated.
To find the energy spectrum for finite wires we use a quantization and discretization scheme that takes $ kx \rightarrow -i\partial/\partial x $ and  ${\partial ^2}c(x)/\partial {x^2} \approx ({c_{i + 1}} + {c_{i - 1}} - 2{c_i})/{b^2}$
and $\partial c(x)/\partial x \approx ({c_{i + 1}} - {c_{i - 1}})/2b$, where $ b $ is the effective lattice constant shown in Fig. \ref{cross_section}. In the following calculations we will set $b$ to be 5 nm.

With periodic boundary condition, that is $ c_{N+1} = c_1 $, there are no Majorana zero modes, and only Andreev states ( shown as Fig. \ref{andreev} ).
Fig. \ref{andreev} (a)-(c) is the density of states(DOS) of the sub-bands with quantum number of (n,m)=(1,0),(1,$\pm1$) and (2,0). We reiterate that the sub-bands with quantum number (1,$\pm1$) are weakly coupled through coupling to other sub-bands, as previously discussed. 
The DOS when Fermi level is tuned at the bottom of the lowest sub-band (that with quantum number of (n,m)=(1,0)) is shown in Fig. \ref{andreev} (a) with the Zeeman energy varying from $ 0 $ to $ 4\Delta_{SC} $ and the edges of the superconducting gap labeled with red arrows. 
When the magnetic field increases, a pair of Andreev states cross when the Zeeman energy equals the pairing potential.
When the periodic boundary condition is removed, that is for finite length quantum wires, the pair of Andreev states evolve into Majorana zero-modes, which remain at zero energy once the Zeeman energy exceeds the pairing potential as in Fig. \ref{andreev} (d).

\ifpdf
\begin{figure}[htbp]
\centering
\includegraphics[width= 1\columnwidth]{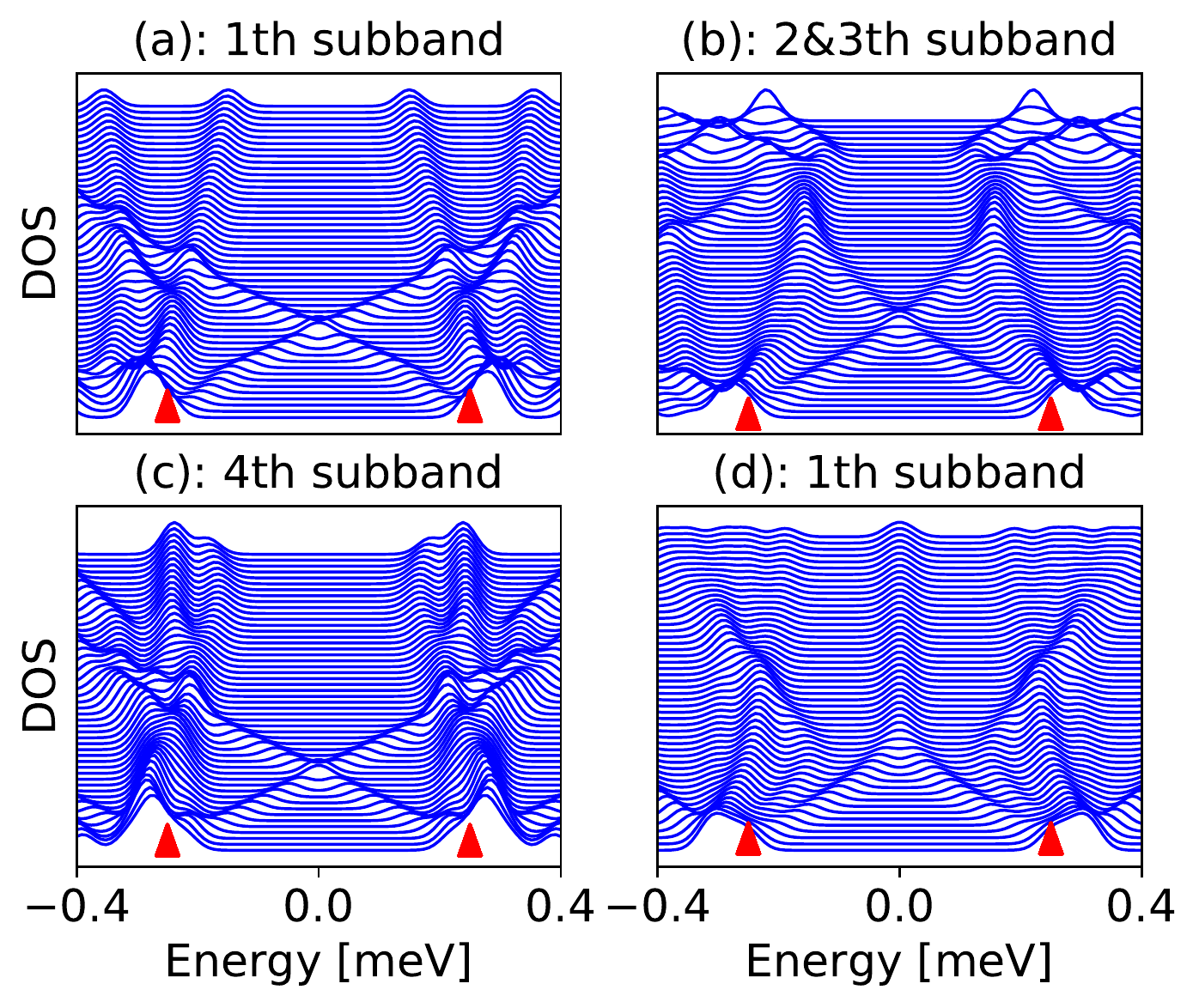}
  \caption{(Color online).
    Envolution of Andreev states and Majorana zero modes versus magnetic field. (a)-(c) show the DOS of the infinite wire when the Fermi level is tuned at the bottom of sub-bands with different quantum numbers. The magnetic field changes from $ 0 $ to $ B = 4\Delta_{SC} $ where $ \Delta_{SC} = 0.25 meV $. The lines are separated by Zeeman energy of 0.02 meV. 
    (d) DOS for finite quantum wire with Fermi level tuned to the bottom of the (n,m) = (1,0) sub-band. The pair of Andreev states evolve into the Majorana zero modes when the Zeeman energy exceed the pairing potential in this case.
}
  \label{andreev}
\end{figure}
\fi

It has been argued that the zero-bias peak observed in experiment can also be caused by disorder.\cite{Disorder_Dmitry,Disorder_Liu,Disorder_Pikulin} Here we construct a binary disorder model for the chemical potential and pairing potential
and use a Gaussian distribution model of the pairing phase disorder.
When the Fermi level lies at the lowest sub-bands, that with (n,m) = (1,0), the DOS of the infinite wire with different kind of disorder are shown in Fig. \ref{andreevDis} (a)-(c). 
To model a charge disorder in the semiconductor we define a spatially varying chemical potential $ \mu_i = \mu \pm \delta \mu $ where  $\delta \mu = |\Delta_{SC}|$ sampled randomly for each site $i$. We find that the DOS is insensitive to this type of disorder (Fig. \ref{andreevDis} (a)).
However, this is not the case for disorder in the phase and amplitude of the superconducting order-parameter (Fig. \ref{andreevDis}(b) and (c)). The disorder is modeled by setting  $ |\Delta_{SC,i}|$ to be 0 and $|\Delta_{SC}| $ randomly for disorder in the pairing potential. Disorder in the phase is model as:
\begin{equation}
    \Delta _{SC,i} = |\Delta _{SC}|e^{i{(\phi_0 + \delta \phi_i})},
\end{equation} 

\noindent
where $ \phi _0 $ is the average phase of the pairing, and the statistics of $ \delta {\phi _i} $ are sampled from a Gaussian function:
\begin{equation}
    f(\delta {\phi _i}) = (1/\sqrt {2\pi } \sigma ){e^{ - \delta \phi _i^2/2{\sigma ^2}}}.
\end{equation}
\noindent
We set the variance of the phase to be bounded by $ \sigma = \pi /2 $.
Disorder in the amplitude and phase lead to substantial changes of the superconductivity gap. From these simulations we see that if we only consider the proximity effect, the experimental finding that the density of in-gap states is much smaller the DOS at the edge of superconducting gap cannot be explained. This suggests that electrons from the superconducting metal may contribute significantly to the experimental DOS.

\ifpdf
\begin{figure}[htbp]
\centering
\includegraphics[width= 1\columnwidth]{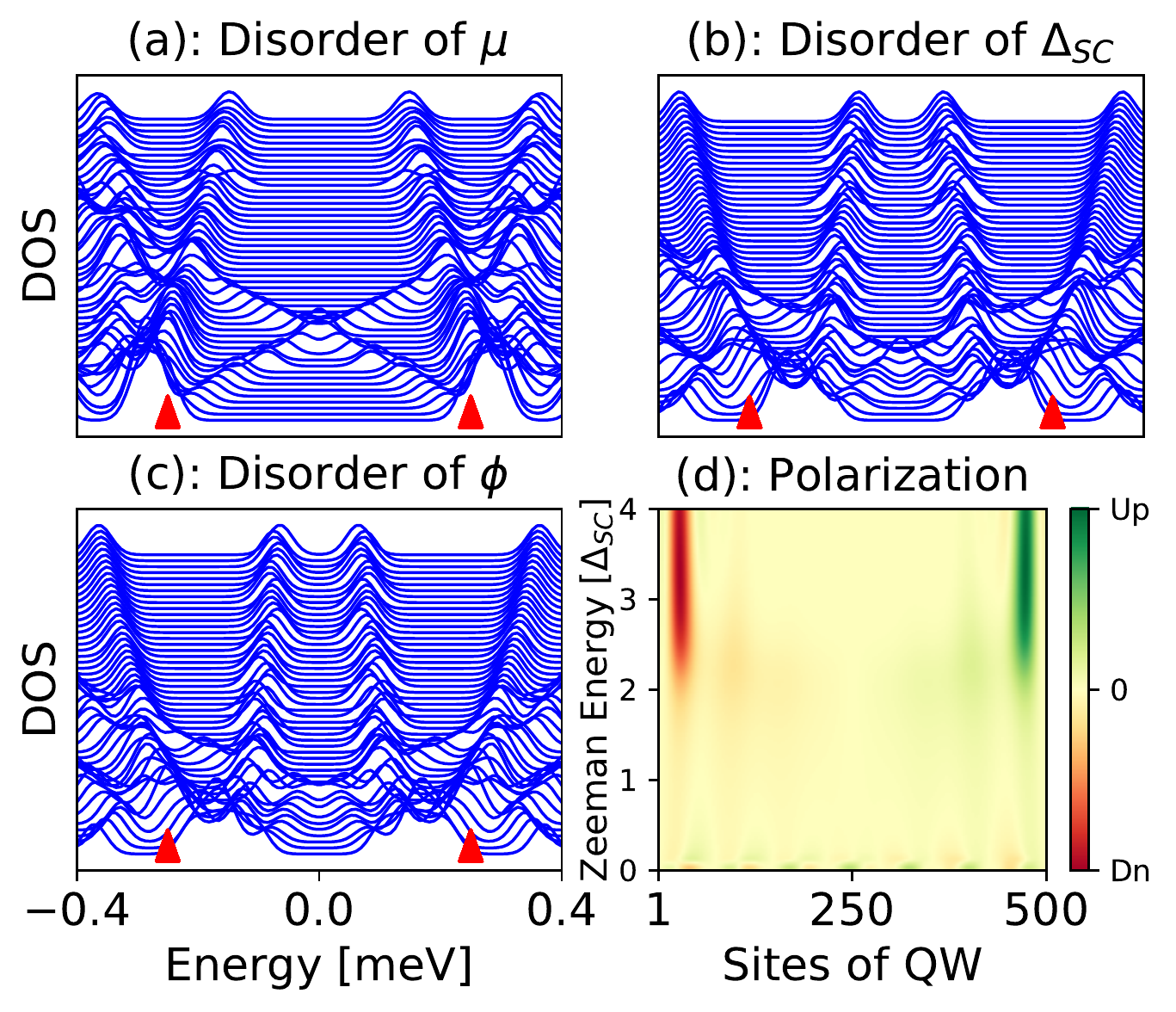}
  \caption{(Color online).
     The effect of disorder and magnetic field on Andreev states.
     (a)-(c) DOS of the infinite wire at the lowest band with different types of disorder:
     (a) disorder in the chemical potential with $ \mu_i = \mu \pm \delta \mu $ where $ \delta \mu = 0, \pm \Delta_{SC} $ by random;
     (b) pairing amplitude disorder with $ \Delta_{SC}^i = 0,\Delta_{SC} $;
     (c) pairing phase disorder modeled as a normal distribution, with the mean phase as $ 0 $ and the variance of the phase as $ \pi /2 $;
     (d) The polarization $ n_{\uparrow} - n_{\downarrow} $ of Andreev states along the quantum wire (here $L = 1\ \mu m$ with periodic boundary conditions).
}
  \label{andreevDis}
\end{figure}
\fi

%NOTE - Chao and Guru will restart from here.

\section{Majorana zero modes properties}\label{majorana_states_properties}

In the previous section, it was shown that disorder in the pairing potential decreases the size of the superconductor gap. We now extend this discussion -- using the same model and model parameters\cite{Expt_Delft} -- now used to describe a finite quantum wire with length of $1\ \mu m$. Fig. \ref{majorana} (a)-(c) shows the DOS versus the Zeeman energy for disorder in the chemical potential, and superconducting order-parameter amplitude, and phase, respectively. The superconductivity gap is robust to disorder in the chemical potential but again decreases with disorder in the superconducting order-parameter amplitude and phase. We find that the critical magnetic field decreases by almost a factor of 2 in the finite wire when disorder in the order-parameter is included.

This sensitivity to magnetic field found in the finite system is not only a symptom of disorder. The Majorana modes in the finite system are more sensitive to magnetic field than the Andreev states seen in the infinite system even in the absence of disorder. The polarization of the two states closest to zero energy, calculated as $n_{\uparrow} - n_{\downarrow}$ in Fig. \ref{majorana}(d) show that the Zeeman energy needed to polarize the Majorana modes is the size of the pairing potential. This can be compared to Fig. \ref{andreevDis}(d) showing the same quantity in the infinite wire where the Zeeman energy required to polarize these states is a factor of $\approx 2$ times the pairing potential. With these results we find that the Andreev bound states in the infinite wire require a larger Zeeman energy to polarize that the Majorana zero-modes even in the absence of disorder.

\ifpdf
\begin{figure}[htbp]
\centering
\includegraphics[width= 1\columnwidth]{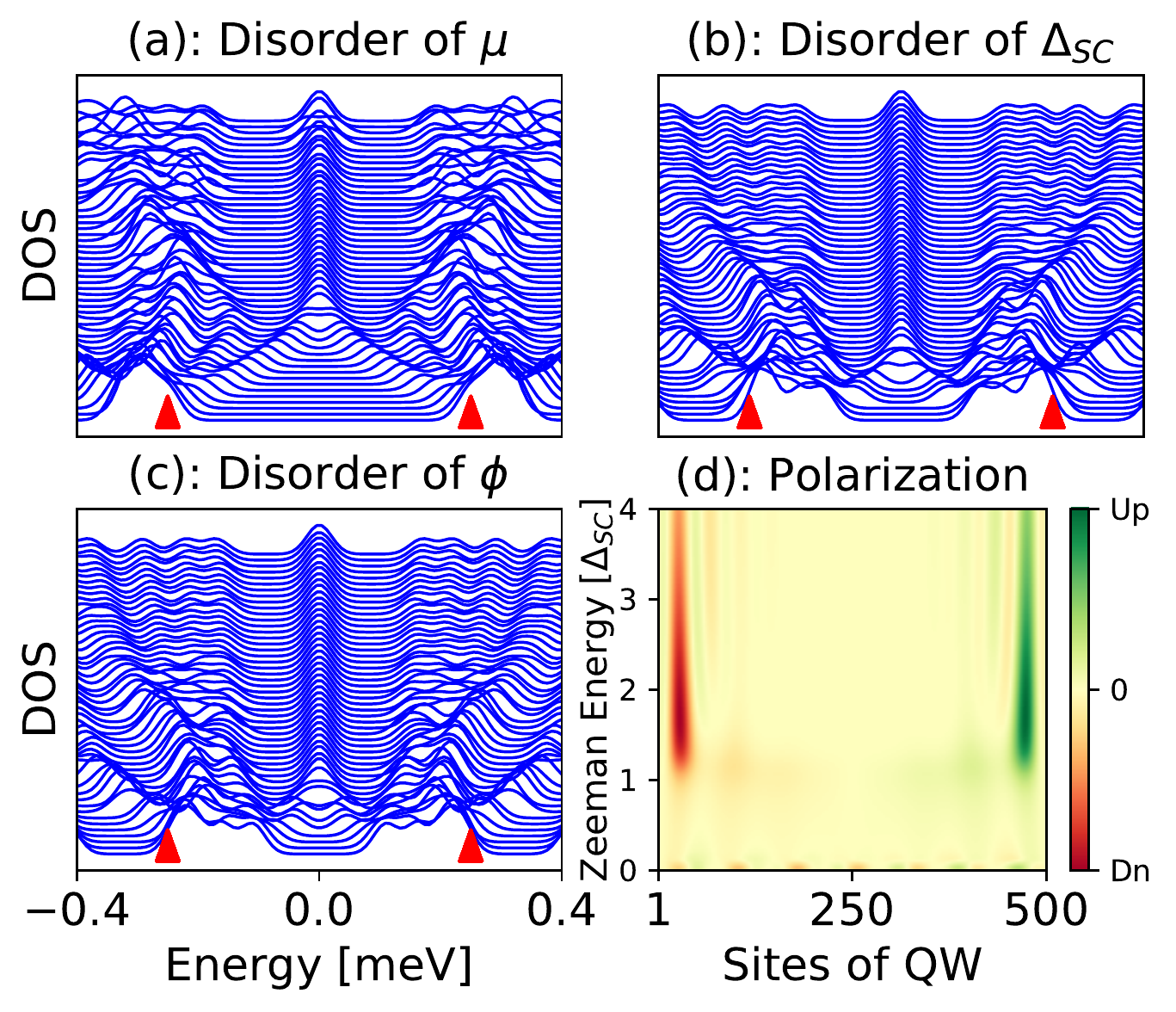}
  \caption{(Color online).
    DOS at the lowest sub-band with quantum number (n,m) = (1,0) in the finite quantum wire. The magnetic field changes from $ 0 $ at the bottom of the plots in increments of 0.02 meV up to $ B = 4\Delta_{SC} $ at the top of the plots. Again, the pairing potential is $ \Delta_{SC} = 0.25 meV $.
    (a) Disorder in the chemical potential;
    (b) Disorder of pairing amplitude;
    (c) Disorder of pairing phase;
    (d) Polarization of Majorana zero modes along the quantum wire (with $L= 1\ \mu m$).
}\label{majorana}
\end{figure}
\fi

%\textcolor{red}{We need to check "pair-potential" and "gap" throughout the manuscript to make sure these quantities are being described correctly. Check the length-scales in the calculations and make sure they are consistent with the values cited in the text.}

%\textcolor{red}{Guru: This study can be used to understand hexagonal cross-section wires and motivated as such. To do so, we reference the appropriate literature in the introduction, and state that we use a cylindrical wire for simplicity but that our results are insensitive to cross-sectional geometry. Then we reiterate this point in the conclusion, and update the abstract.}

%\textcolor{red}{Guru: The justification for the perturbed superconductor needs to be made a bit stronger -- at least in the Discussion section.}

In tunneling experiments on proximitized quantum wire systems, the local density of states at the end of the quantum wire is probed. The DOS at the edge of the superconducting gap is found to be larger than the zero-biased peak associated with Majorana modes. We can probe this feature in our model by calculating the projected DOS for different length-scales measured from the end of the wire. Fig. \ref{pdos} (a) and (b) shows the projected DOS parameterized by the Zeeman energy for states projected within 50 nm, and 250 nm from the end of the wire. We note again that the length of the wire in this simulation is 1 $\mu\ m$. In this way we can compare states at the end of the wire with the bulk system. While we find the DOS near the edge of the gap is small when we focus on the end of the wire (Fig. \ref{pdos} (a)), this become comparable with the zero-bias peak when we include bulk states (Fig. \ref{pdos} (b)). The size of the DOS at the superconducting gap energy relative to the zero-bias peak found in experiment, cannot be accounted for in this model. 

\ifpdf
\begin{figure}[htbp]
\centering
\includegraphics[width= 1\columnwidth]{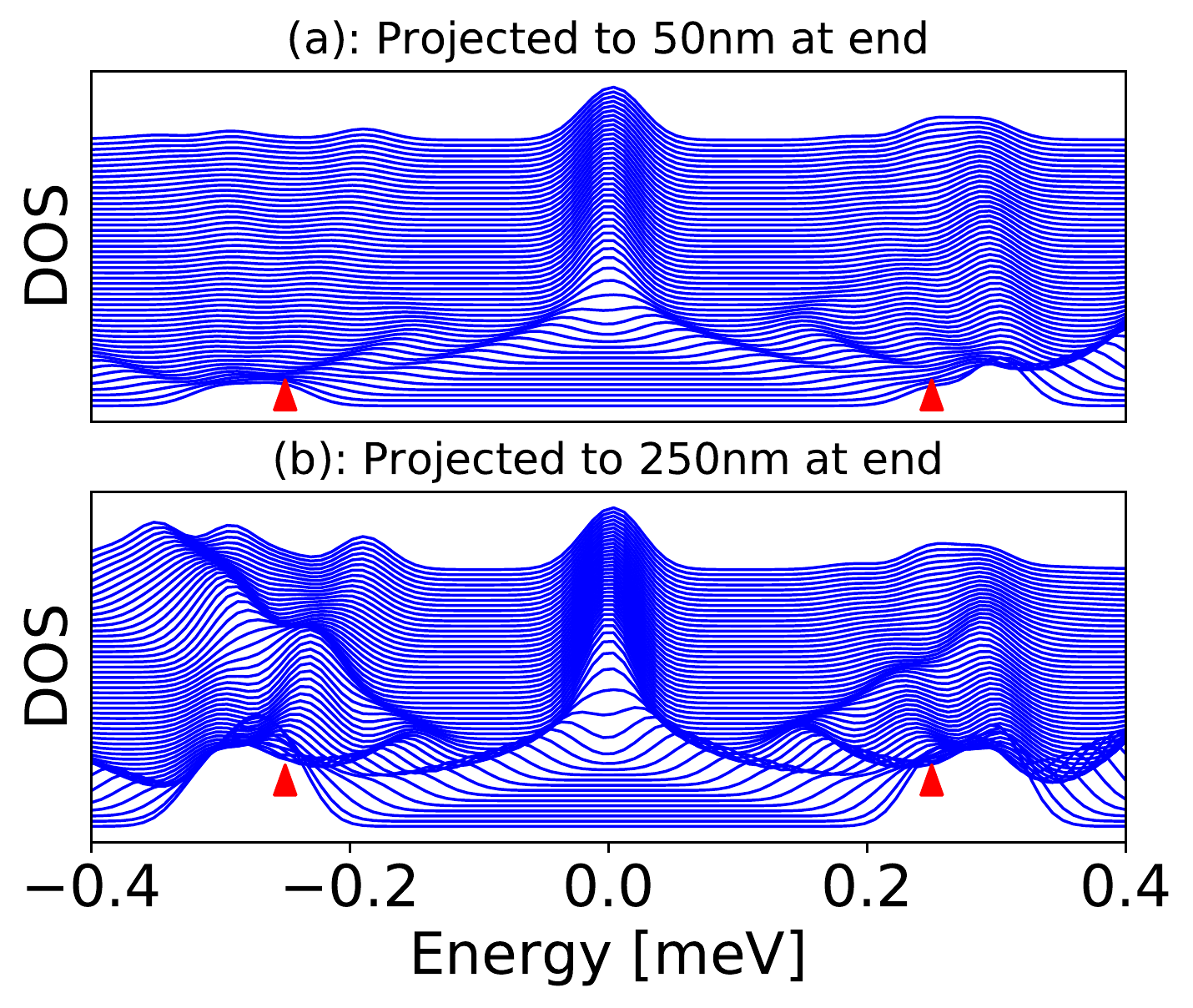}
  \caption{(Color online).
    Projected DOS at the end of the quantum wire.
    (a) is the results of DOS projected at the end within $ 50 nm $ and
    (b) is the results of DOS projected at the end within $ 250 nm $.
}
  \label{pdos}
\end{figure}
\fi

\section{Discussion}\label{discussions}

%In this paper we modeled semiconductor quantum wires with cylindrical cross-section, discussed the Andreev states and Majorana zero modes in quantum wire with present experimental parameters and studied the influeces of various disorder, we also showed the polarization of Andreev states and Majorana zero modes and the projected DOS at the end of semiconductor quantum wire. 

In this paper we have studied the properties of cylindrical semiconducting quantum wires proximity coupled to a superconductor. 
Topological states occur in the presence of an external magnetic field for Fermi levels just above the population thresholds of all angular momentum $m=0$ quasi-one-dimensional sub-bands.
For long but finite cylindrical wires, Majorana zero modes are localized near wire-ends at $m=0$ sub-band population thresholds.  
In contrast, pairs of localized Majorana-like states appear at each end near the populations thresholds of degenerated sub-bands which can give rise to zero-bias anomalies in transport -- although they are not strictly speaking associated with topological states.

By modeling the influence of disorder within quantum wires we conclude that if only the proximity effect in semiconductor quantum wire is considered, the only effect of pairing(or phase) potential is to give a smaller observable superconducting gap, disorder within the quantum wires is thus not on its own able to account for the discrepancies between naive quantum-wire theory and experimental findings.
We suggest instead that the current-path followed in the transport experiments used to probe semiconductor quantum wires is strongly influenced by sub-gap states in the superconducting metal which surrounds the quantum wire and is used to induce superconductivity within it.

%\textcolor{red}{
%Move the derivation of the number of electrons in the semiconductor and superconductor to an appendix. Discuss the failures of the model in the discussion section and reference the scale of the ratio of the number of electrons from the Appendix. The semiconductor model, even when compared with local probes at the edge of the wire, should only be used to account for features within the gap and near the zero-bias peak. Suggest a model for future work. Do not include an effective model in the Appendix. 
%}

The DOS measured in experiment strongly depends on not only the Majorana zero modes spectrum, but also the superconducting element which donates its superconductivity to the semiconductor quantum wire. We clarify this point by estimating the semiconductor and metal (superconductor) electron density for direct comparison (See Appendix \ref{summary_theory_appendix} for details).

We find that the number of electrons in the metal/superconductor $N_{sc}$ greatly outnumbers the number of electrons in the semiconductor $N_{qw}$ ($N_{sc} >> N_{qw}$), with their ratio ranging from $ \sim 10^3 $ to $ \sim 10^5 $ depending on the specific materials. 
Summarized in Table \ref{summary_exp_qwsc}, are the estimates of the electron count. The number of electrons in the superconducting metal is many orders of magnitude larger than the number of electrons in the semiconductor quantum wire.

\begin{table}[h]
\caption{\label{summary_exp_qwsc}
Extracted experimental parameters. Ratio of the superconducting pairing potential $\Delta$ to the spin-orbit energy $E_{so}$, the number of electrons in the quantum wire $N_{qw}$, and the ratio of the number of electrons in the superconductor $N_{sc}$ to the number of electrons in the quantum wire. 
}
\begin{ruledtabular}
\begin{tabular}{l c c c }
 Materials & $ \Delta/E_{so}$ & $N_{qw}$ & $ N_{sc}/N_{qw} $ \\ [0.5ex]
  InSb/Nb\cite{Expt_Delft} &  $  0.8  $ & 5.0 & $  2.0\times10^5 $  \\
  InSb/Nb\cite{Expt_Rokhinson} & $  0.6  $ & 1.4 & $  2.3\times10^4 $ \\
  InAs/Al\cite{Expt_Weizmann} & $  0.8 $ & 0.4 & $  1.3\times10^4 $ \\
  Nb/InSb/Nb\cite{Expt_Xu} & $  0.3 $ & 3.0 & $  1.4\times10^5 $  \\
  InSb/NbTiN\cite{Expt_Marcus} & $  1.4  $ & 0.6 & $  2.3\times10^5 $ \\
  InAs/NbN\cite{Expt_Finck} & $  16.5 $ & 1.9 & $  4.8\times10^4 $ \\
  InAs/Al\cite{albrecht2016,Deng2016,Zhang2018} & $  2.1 $ & 0.6-2.5 & $ 9.3\times10^2 $ \\
 \end{tabular}
 \end{ruledtabular}
\end{table}

This shows that electrons in the superconducting metals will play an important role in understanding measurements of Majorana zero-modes in proximatized quantum wires. A complete model that include both electrons in semiconductor quantum wires and superconducting metal \cite{ReviewStanescu} is thus necessary.

The present proximitized semiconductor quantum wire based Majorana systems may be in fact treated as a superconducting metal perturbed by magnetic field and spin-orbit interaction proximitized by semiconductor quantum wires, the main contribution to the tunneling DOS come from the electrons in superconducting metal instead of the semiconductor quantum wire. Ultrathin film metals with strong spin-orbit coupling\cite{metal_majorana} are thus a prospective platform to realize topological superconductors if the g-factor is large enough and effective tools are found to tune the Fermi level.

\section{Acknowledgements}  This work was financially supported by Welch Foundation under grant TBF1473 and by Army Research Office under Grant Number W911NF-16-1-0472.

\appendix

\section{Estimation of experimental parameters}\label{summary_theory_appendix}
To estimate the electrons envolved in Majorana zero modes in semiconductor quantum wire,
we consider the active sub-band and model it with the follow quasi-one-dimensional Hamiltonian:
\begin{equation}
 H_k = \frac{\hbar^2}{2m^{\ast}} k^2 + \alpha k \sigma_y
\end{equation}
where $\hbar$ is the reduced Plank constant, $m^{\ast}$ is the effective mass of electrons in semiconductor, $\alpha$ is the Rashba coupling.
The band energy can be solved to be:
\begin{equation}
 E_k = \frac{\hbar^2}{2m^{\ast}} k^2 \pm \alpha k.
\end{equation}
To estimate the number of electron in the semiconductor quantum wires $N_{qw}$, we take advantage of the quasi-1D nature of the wires and find $N_{qw} = \frac{k_{so}\cdot L}{\pi}$(see Table \ref{summary_exp_qwsc}). 
Here $k_{so} = \frac{2\alpha m^{\ast}}{\hbar^2}$ is the spin-orbit wave vector and L is the length of quantum wire. This assumes that the chemical potential has been tuned to $E_{so}$ by the gate voltage.
The spin-orbit wave vector and spin-orbit energy $E_{so} = \frac{\alpha^2 m^{\ast}}{2\hbar^2}  = \frac{\alpha}{4}k_{so}$ are estimated from the extracted experimental effective electron mass $m^{\ast}$ and Rashba coupling $\alpha$. The estimation of the experimental parameters in semiconductor quantum wire are shown in Table \ref{summary_exp_qw}.

\begin{table*}[htp]
\caption{\label{summary_exp_qw}
Summary of parameters of semiconductor quantum wires.
}
\begin{ruledtabular}
\begin{tabular}{c c c c c c c c c c c c c c c }
 Materials & Geometry & L[nm] & $\alpha[eVnm]$ & $m^{\ast}[m_e]$ & $k_{so}[nm^{-1}]$ & $ \lambda_{F}[nm] $ & $r_{ee}[nm]$ & $ E_{so}[meV] $ & $ N_{qw} $  \\ [0.5ex]
  InSb/Nb\cite{Expt_Delft} & Cir & $\sim 2000 $ & 0.02 & 0.015 & 0.0079 & 127 & 399 & 0.315 & 5 \\
  InSb/Nb\cite{Expt_Rokhinson} & Ret &  $\sim 600$  & 0.019 & 0.015 & 0.0075 & 134 & 420 & 0.284 & 1.4 \\
  InAs/Al\cite{Expt_Weizmann} & Cir &  $\sim 150 $  & 0.0113 & 0.03 & 0.0089 & 112 & 353 & 0.201 & 0.4  \\
  Nb/InSb/Nb\cite{Expt_Xu} & Cir & $\sim 740(680) $ & 0.032 & 0.015 & 0.0126 & 79 & 250 & 0.806 & 3(2.7) \\
  InSb/NbTiN\cite{Expt_Marcus} & Cir & $\sim 250 $ & 0.02 & 0.015 & 0.0079 & 127 & 399 & 0.315 & 0.6 \\
  InAs/NbN\cite{Expt_Finck} & Cir & $\sim 1000 $ & 0.01 & 0.023 & 0.006 & 166 & 520 & 0.121 & 1.9 \\
  InAs/Al\cite{albrecht2016,Deng2016,Zhang2018} & Hex & $330-1500 $ & 0.008 & 0.025 & 0.0052 & 190 & 598 & 0.084 & 0.6-2.5 \\
 \end{tabular}
 \end{ruledtabular}
\end{table*}

Via proximy effect, the Cooper pairs tunnel into the quantum wire, the DOS in Aluminum is
\begin{equation}
 D(E_F) = \frac{m^\ast}{\pi^2\hbar^3}\sqrt{2m^\ast E_F} = \frac{2m^\ast}{\hbar^2 k_F^2}\frac{\hbar^2 k_F^2}{2\pi^2}\sqrt{\frac{2m^\ast}{\hbar^2} E_F}
\end{equation}
since $E_F = \frac{\hbar^2 k_F^2}{2m^\ast}$, then
\begin{equation}
 D(E_F) = \frac{1}{E_F}\frac{k_F^2}{2\pi^2}\sqrt{\frac{1}{E_F}k_F^2 E_F} = \frac{1}{E_F}\frac{k_F^3}{2\pi^2}
\end{equation}
while the density of free electron in 3D system is $n = \frac{2\cdot 4\pi k_F^3}{(2\pi)^3} \to k_F^3 = 3\pi^2 n$, then the DOS is
\begin{equation}
 D(E_F) = \frac{3n}{2E_F}
\end{equation}
To calculate the number density of free electrons (n):
\begin{equation}
 n = z\frac{N_A}{V_A}
\end{equation}
where $z$ is the valency,$N_A$ is the Avogadro’s constant,$V_A$ is the molar volume.To calculate the molar volume:
\begin{equation}
 V_A = \frac{M_r \times 10^{-3}}{\rho}
\end{equation}
where $M_r$ is the relative atomic mass (the $10^{-3}$ is to convert $M_r$ from grams to $kg$),$\rho$ is the density. We then get
\begin{equation}
 n = \frac{z \rho N_A}{M_r \times 10^{-3}}
\end{equation}
For Aluminum $z = 3$ and $M_r = 27$, while for Niobium, $z = 5$ and $M_r = 93$, and the Avogadro constant is $6.02\times 10^{23}$, then
$ n = 1.8\times 10^{29} m^{-3} $ for Aluminum and $ n = 2.8\times 10^{29} m^{-3} $ for Niobium.

The number of electron in the superconducting metal is estimated
by:
\begin{equation}
 N_{so} = D(E_F)\cdot E_{so}\cdot V_{sc} = \frac{3n}{2}\frac{E_{so}}{E_F}\cdot V_{sc},
\end{equation}
where $V_{sc}$ is the volume of the superconducting shell, this expression for $N_{sc}$ assumes that the DOS is constant on the scale of $E_{so}$ and that only electrons near $E_{so}$ contribute, the corresponding parameters estimated from experiments are shown in Tabel \ref{summary_exp_sc}.

\begin{table}[b]
\caption{\label{summary_exp_sc}
Summary of parameters of superconducting metals.
}
\begin{ruledtabular}
\begin{tabular}{c c c c }
 Materials & $V_{SC}[10^6nm^3]$ & $ \Delta [meV]$ & $ N_{sc}[10^4] $ \\ [0.5ex]
  InSb/Nb\cite{Expt_Delft} & 40 & 0.25  & 99.46  \\
  InSb/Nb\cite{Expt_Rokhinson} & 1.44 & 0.18 & 3.23  \\
  InAs/Al\cite{Expt_Weizmann} & 1.18 & 0.15 & 0.55  \\
  Nb/InSb/Nb\cite{Expt_Xu} & 6.65(6.11) & 0.25 & 42.33(38.89)  \\
  InSb/NbTiN\cite{Expt_Marcus} & 5.89 & 0.45 & 14.65 \\
  InAs/NbN\cite{Expt_Finck} & 9.64 & 2 & 9.19  \\
  InAs/Al\cite{albrecht2016,Deng2016,Zhang2018} & 0.264-1.2 & 0.18 & 0.05-0.23  \\
 \end{tabular}
 \end{ruledtabular}
\end{table}

\bibliography{SemiMFref}

\end{document}